\documentclass[11pt]{article}
\usepackage{graphicx}
\usepackage{amsmath}
\usepackage{geometry}
\usepackage{amsfonts}
\usepackage{amssymb}
\newtheorem{theorem}{Theorem}

\newtheorem{remark}[theorem]{Remark}

\begin{document}

\author{{\Large Scott M. Hitchcock}\\National Superconducting Cyclotron Laboratory (NSCL)\\Michigan State University, East Lansing, MI 48824-1321\\E-mail: hitchcock@nscl.msu.edu}
\title{{\Huge Time and Information}}
\date{February 4, 2001}
\maketitle
\begin{abstract}
The relationship between 'information' and 'time' is explored in order to find
a solution to the 'Problem of Time'. 'Time' is a 'result' of the conversion of
energy into 'information'. The numbers or labels assigned we give to the
'times' when events occur can be '\emph{computed'} from the processing of
state information 'flowing' from a Feynman Clock (FC) \cite{hitchcock},
\cite{hitchcock2}, \cite{hitchcock3}, via a 'Signal', to a Feynman Detector
(FD). The ordered set of the 'time' numbers labeling these 'events' can be
used to construct the 'direction' and 'dimension' associated with the usual
conception of a 'time' axis or coordinate in quantum, classical and
relativistic mechanics.
\end{abstract}

\section{Introduction}

\begin{quotation}
\textbf{What, then, is time? If no one asks me, I know what it is. If I wish
to explain what it is to him who asks me, I do not know}-\emph{St. Augustine
\cite{harrison}}
\end{quotation}

\noindent Can we discover what 'time' is by looking at it as 'information'
created by the flow of energy between objects in the universe? If so how is
the 'time' produced from this 'information'? What do we mean by 'information'
in the context of the quantum systems that compose our world? 'Information'
has many meanings from the \emph{description }of the quantum state of an atom
in an atomic clock to complex abstract\emph{\ ideas }about the entire atomic
clock 'system' in the 'imagination' of the brain.

\begin{quotation}
\textbf{Information is Physical}-\emph{Rolf Landauer\cite{qcqi}}
\end{quotation}

\textbf{Information }is treated as a \textbf{physical} aspect of our world. It
is more than just labels, words, or 'bits' since the \emph{physical state} of
a system contains the 'information' which will be processed or computed into
the 'times' we use to label the order of events in causal networks and
sequences. The quantum world alone is not 'timeless' but has irreversible
properties when considering excited states. The quantum world can be
'reversed' or more properly 'reset' by interacting with signals and energy
from its 'classical' environment. More specifically 'time' is rooted in the
states of systems that create signals which in turn transfer information to
other systems. We will define primary state information to be a physical
observable or measurable state of a system that can be coupled, paired or
'entangled' with another signal such as a clock generated pulse. A 'triplet'
state in a 'memory' or detector can be formed consisting of the signal induced
state, the clock pulse state, and the time label counter state that tags the
clock pulse with a label. The creation of a triplet states involves a process
of 'signal mapping' which will be discussed later.

The triplet state can be operated on to extract the 'time' in the form of
\textbf{secondary state information}. These numbers are the 'times' that the
observer (quantum information processing system) pairs with the observed
signals from the 'source' events. The real numbers from the ordered set of
events can be used to from a 'dimension' or axis of 'time' mapping the events
to a standard clock. The 'dimension of time' is just the dimension of the real
number 'line' used to map the 'event time'. Further 'processing' of
information mapped on 'timelines' represents \textbf{tertiary state
information.} This tertiary information is exemplified in higher level
mappings of the 'times' to the mathematical and other 'abstract' descriptive
languages used to establish to 'meaning' of the temporal information in models
of cause and effect for the behaviors of complex systems. Since conventional
'time' is a derived measure of information flow through evolving systems, it
should come as no surprise that its application in theoretical models can take
many forms. 'Imaginary' and 'multi-dimensional' times are 'possible' in that
maps of phenomena can be created using them but these 'constructions' probably
ascribe a 'temporal' property to some poorly understood 'non-temporal' aspect
of the world that is fundamentally 'informational'.

\section{Information, Clocks, and Causal Networks}

\begin{quotation}
\textbf{Time must never be thought of as pre-existing in any sense; it is a
manufactured quantity}-\emph{Hermann Bondi\cite{davies}}
\end{quotation}

Where does this information originate? Perhaps we can see the essence of the
'problem of time' by comparing two identical 'working' electronic watches. One
with a battery and one without. What is the difference between these watches?
Energy. The 'powered' watch is an example of a state repeating cyclical
'clock' that generates time 'information' by conversion of the energy
(supplied by the battery) into 'signals' that can be read or detected by an
observing system or person. The battery provides energy and therefore
re-configuration information from the watches '\emph{environment}'. The
internal configuration of the atoms and molecules of the watch 'ratchet 'or
'gate' the energy flow through the mechanism creating an excited state (tick)
which decays to a ground state (tock). This conversion of energy into
information is perceived as either the 'movement of hands' or the display of
sequential numbers. These signals define the 'time' labels we map to other
observed events. The working watch 'manufactures' the signals that we process
and interpret as 'time'. The watch without the battery is 'timeless' providing
no other forces act upon it to change its structure.

\begin{quotation}
\textbf{It would be quite appealing if atoms interacting with photons (or
unstable elementary particles) already carried the arrow of time that
expresses the global evolution of nature.-\emph{Ilya Prigogine and Isabelle
Stengers }}\emph{\cite{order}}
\end{quotation}

Let us look at a quantum watch whose energy to information conversion occurs
at the quantum scale. These quantum watches can be created as one-time-only
instant clocks at the site of multiple particle collisions in a target or they
may be quantum watches waiting for 'batteries' in the form of photons for
instance as in the case of atoms in their ground states. Once an unstable
configuration of matter and energy is created in a local region of space, we
can say that a watch or clock now exists there. In the case of quantum
instabilities in space, the total process mapping the incoming to the outgoing
signals can be modeled from Feynman Diagrams in which 'time' is removed as a
'dimension' since the creation of unstable excited states occurs in space but
not 'in time'. A\emph{\ time-independent} Feynman Diagram is used to map the
types, energies and '\textbf{trajectories}' of 'incoming' fundamental
particles resulting in the creation of a \textbf{Feynman Clock} (or
\textbf{FC}).\emph{\ }

\begin{remark}
\textbf{Note that a 'trajectory' in this paper is really a composite 'system'
(SEN) composed of a source (FC, CEN), signal (e.g. photon, phonon, etc.),
detector (FD mode of a FC, or a CEN) and a \emph{spatial map} of these quantum objects.}
\end{remark}

\subsection{Three Assumptions}

We will make a conceptual leap inspired by the quantum hypothesis which
assumed that 'particles' can be here then there without passing through the
space in between. In the following theory of 'time' as a form of 'information'
we will make \emph{three assumptions}.

The \emph{first assumption} is\emph{\ }that \emph{the method of Feynman
Diagrams }\cite{Veltman}, \cite{Mattuck}, used to understand particle
collisions or 'reactions' at extremely high energies, can be modified into
\textbf{'time-independent}\emph{' maps} (FCs), that represent the
\emph{initial} and \emph{final} \textbf{configuration states} of a single
'system'. This system may be formed by incoming particles interacting with a
target or '\textbf{quantum clock}' \cite{Peres}. The 'transitions' from the
initial excited state to any of many possible 'final' lower energy states is
accompanied by the production of outgoing signals. These configuration
transformations are 'time-independent' in the sense that the transitions are
fundamentally \textbf{geometric}. Any conventional '\emph{times}' associated
with these \emph{events} are 'labels' that identify the causal and therefore
'temporal' relationships of the states of the changing system in the
observer's detectors (e.g. eyes). In other words the total 'geometric'
configuration (as identified by the mass and energy distributions of the
components of the system in space) are in a sense 'quantized' into
configuration states when the system acts as a single entity. This allows us
to look at these special configuration transitions, in an evolutionary
universe, not as happening '\emph{in time}' but producing 'time' information
carried by signals created in the \emph{decay process}. Once the signals (e.g.
photons, phonons, images, etc.) are 'detected' they can be labeled or stamped
with 'time numbers' by the observing system. This process couples or entangles
the induced information states in a detector (or 'memory') with 'coincident'
standard clock states. The classical 'event time' can be extracted by
disentangling this complex compound state into a classical time 'number'.
Ordered sets of these number and event pairs can be used to create 'timelines'
or the time 'axis' of \emph{space-time}.

The \emph{second assumption} is that once an unstable state or configuration
of matter is created in a quantum system, it must then decay irreversibly.
During this process 'signals' are created. The FC decays with the production
of outgoing 'signals' and perhaps leaving a remnant (in the center of mass
frame) particle or \textbf{Feynman Detector} (or \textbf{FD}). It is at the
point of creation of a FC that the\textbf{\ irreversible quantum arrow of
time} \textbf{(QAT)}can be defined with respect to a lower energy
reconfiguration state or FD. The quantum arrow of time is not really a 'time'
arrow but a one-way function mapping the unstable configuration of a system to
a more stable one \cite{Diu}. The QAT represents 'temporal directionality' at
the most fundamental scale. It has been overlooked because of the confusion
between 'time reversal' and 'process reversal' in microscopic physics.

This system could still be reconfigured into a FC state by detection of
another signal. This \textbf{is not} '\emph{time reversal}' but a
time-independent '\emph{process reversal}' in which a 'signal' from the
classical environment reconfigures a FD into a FC. The 'problem of time' is
the 'paradox' of the apparent \emph{reversibility} 'in time' of particle
collisions compared to the apparent irreversibility of the macroscopic world
built from these quantum systems. This is characterized by the Second Law of
Thermodynamics with its associated 'entropy' directed 'arrows of time'. The
key to the solution of the 'problem' is that the reversibility of the particle
collision process \emph{does not imply} that the unstable state of the
transient aggregate quantum system (FC) is 'reversible', since it is the
unstable state that must decay irreversibly!

The \emph{third assumption} is that networks of FCs (acting 'internally' with
their irreversible QATs and driven by detection or processing of external
\textbf{reversible} \textbf{Classical Arrow of Time (CAT)} 'reset' or
excitation signals) can support \textbf{collective excitations states}
\textbf{in networks} that \textbf{decay irreversibly}. This is the basis for
\textbf{Collective Excitation Networks (CENs)} \ and \textbf{Collective
Excitation (CE)} signals such as those seen in nuclear multipole vibrations,
condensed matter interactions with phonons and 'sound waves', global quantum
states in superconductors, and excitons in photosynthetic
'detectors'\cite{ritz}. Collective excitations are the key to understanding
how hierarchical systems emerge from their microscopic components. The QATs of
these atomic and molecular components acting collectively as a single system
can generate the arrows of time that we associate with chemical, biological
and cosmic processes. The key is that the coupling of these nodes in a network
now form a new larger scale 'quantum' system.

A 'classical' system emerges when it's basic network components
(Signal-'connectors', and the FC or CEN 'nodes' and 'gates') are 'space-like'
separated with respect to information flow. In other words if the states of
the components can't be coupled or '\textbf{entangled'} \cite{zurek},
\cite{zeh} \ in order to support a collective state then the information flow
may be\ a causal sequence of \ 'isolated' states in a step-by-step information
transfer process which we will call a \textbf{Sequential Excitation Network
(SEN}). Examples of this include, 'conventional' computers, sequential
chemical reactions in the cell, nerve impulse transmission, and higher level
metabolic processes in animals. FCs and CENs form the building blocks of SENs
which can also act like a 'single' system with novel collective 'classical' behaviors.

\section{'Sources' of Information}

The reconfiguration or 'decay' of an unstable quantum system (a Feynman Clock
or FC) into a more 'stable' state (Feynman Detector mode of the FC) plus a
'signal', occurs with a finite 'lifetime'. This 'time number' is created by
the 'dimensional' conversion (using Planck's Constant) of the '\emph{energy of
reconfiguration}'\textbf{\ information}, $\mathbf{I}_{R}$ (note that this is
\textbf{primary state information}), into the 'time' associated with the
reconfiguration of the system in an excited state to a lower energy
configuration. The \textbf{mapping} of the reconfiguration information to a
real number is the result of the interaction of the unstable quantum system
with its classical 'environment'. The 'map' between the quantum system and its
environment is Planck's Constant. This \textbf{Planck Mapping Function},
$\hbar$, is the quantitative measure of the coupling of quantum systems to
their 'environments'. This function 'resides' in the 'observer' system and
relates the primary information at the quantum scale to the derived
('computed') secondary information from it. Note that secondary information is
also 'physical' in that it may be the state of a single quantum system or a
collective state in a network of these systems.

\section{The Quantum Arrow of Time}

We start with a simple two-level quantum system (Feynman Clock) with only one
excited and one ground state; $\left|  E^{\ast}\right\rangle \equiv\left|
1\right\rangle $ and $\left|  E_{0}\right\rangle \equiv\left|  0\right\rangle
$ respectively. A photon 'signal' with energy, $\Delta E_{\nu}=E^{\ast}%
-E_{0}=h\nu=hc/\lambda$ is heading towards the FC in state $\left|
0\right\rangle $. The composite 'classical' system $\left|  S_{C}\right\rangle
$ composed of the 'free' photon $\left|  \lambda\right\rangle $ and the target
FC $\left|  0\right\rangle $ is:%

\begin{equation}
\left|  S_{0}\right\rangle =\left|  \lambda\right\rangle +\left|
E_{0}\right\rangle =\left|  1\right\rangle +\left|  0\right\rangle
\end{equation}

Absorption of the photon results in the creation of the excited FC state
represented by direct or tensor product of the photon and target states:%

\begin{equation}
\left|  S^{\ast}\right\rangle =\left|  \lambda\right\rangle \otimes\left|
E_{0}\right\rangle =\left|  1\right\rangle \otimes\left|  0\right\rangle
\Longrightarrow\left|  0\right\rangle \otimes\left|  1\right\rangle =\left|
\lambda\right\rangle \otimes\left|  E^{\ast}\right\rangle =\left|
0,1\right\rangle =\left|  S_{FC}\right\rangle
\end{equation}

This is now a FC in an unstable configuration. It will decay irreversibly with
a 'lifetime' determined by the degree of instability created by the internal
fundamental interactions between the matter and energy distributions of the
components of the system. The 'geometry' of the system is driven by these
interactions into a more 'symmetric' configuration in which the energy of the
total system is minimized. The decay 'lifetime' \cite{Diu} of this excited
state is:%

\begin{equation}
\tau_{S_{FC}}\equiv\frac{\hbar}{\Gamma_{FC}}=\frac{\hbar}{\left|
\langle1,E_{0}\mid H_{E^{\ast}\rightarrow E_{0}}\mid0,E^{\ast}\rangle\right|
^{2}}=\dfrac{\hbar}{\mathbf{I}_{R}}%
\end{equation}

Note that the denominator, $\mathbf{I}_{R}$, is usually referred to as the
'decay rate', $\mathbf{\Gamma}$. The $\left|  \langle\Psi_{(E_{0})}\mid
H_{E^{\ast}\rightarrow E_{0}}\mid\Psi_{(E^{\ast})}\rangle\right|  ^{2}$ term
has units of \ 'energy'. This is the \textbf{reconfiguration information}
encoded in the energy transported by the signal that exits the quantum system
into the classical environment. An \textbf{image state }is created in a
spatially distinct Feynman Detector by resonant absorption of the signal. The
image state marks the transition of the FD into its FC mode. The source FC,
signal, and FD together define a \emph{node-arc-node} segment (a
'\textbf{trajectory}' for 'information' qubits as defined above). Connected
segments can be used to build hierarchical causal networks in which the flow
of 'information' defines a primitive form of 'quantum computer'.

The decay of the excited state is irreversible in the sense that once the
unstable state is created it must decay. The bottom term in the above equation
is the 'intrinsic energy of reconfiguration' for an unstable quantum system.
The system in a ground state will not spontaneously become excited without a
signal from it's environment. The 'excited' state may be \emph{recreated} by
another incoming photon of the same energy, but this requires information to
be put into the FC from it's environment. This is not 'time reversal' but a
'reconfiguration' process.

The excited state transition to a ground state allows us to \emph{create} an
\emph{irreversible} '\textbf{Quantum Arrow of Time'(QAT)} \emph{which always
'points' from the unstable state to a more stable one}. \ The QAT applies to
excited states for individual particles and also systems of particles that act
\emph{collectively} as a single larger 'quantum' system. Collective
Excitations (CEs) of these composite systems are the key to understanding
irreversible processes in larger systems composed of many interacting FCs.
These Collective Excitation Networks (CENs) of FCs can support new behaviors
and states in complex systems such as those seen in giant nuclear multipole
vibrations, phonon scattering and absorption of photons in solids, sound
waves, and electromagnetic waves in the brain.

The QAT defined by the irreversible decay of unstable states of quantum
systems\emph{\ }\textbf{closes} ''\emph{the gap between the dynamical
description of quantum mechanics and the evolutionary description associated
with entropy}''(Ilya Prigogine, P.47, \cite{cert}) and solves the 'quantum
paradox' if we see that the decay of an unstable state is a form of
'self-measurement'. The 'decay' or initiation of the reconfiguration process
of the internal 'geometry' is triggered by fundamental interactions (forces)
acting on the asymmetric distributions of energy and matter. The fundamental
instabilities involved with excited states of matter provide the basis for the
irreversible QAT and therefore create the possibility of a 'unified'
formulation of quantum theory and the fundamental interactions of matter
\cite{hitchcock3} from their origins in the Big Bang to their 'expression' as
the complex collective states of consciousness occurring in the brain.

\section{The Classical Arrows of Time}

The environment of a Feynman Clock can provide information (typically from the
signal created by another clock) in the form of energy carrying resonant
signals that can reconfigure or reset a system into an unstable state. Once a
'signal' in space has 'decoupled' from its' source at the quantum level, one
can think of the signal as mapping information flow from a source to a
detector in a classical space (ignoring for the moment 'interference term's
arising from interactions of entangled signals with other quantum systems).
The classical 'trajectory' of the signal can be thought of as Sequential
Excitation Network of the vacuum between source and the target detector. This
'trajectory' is a \textbf{Classical Arrow of Time (CAT).} The key to the
classical arrow of time is the 'space' in which quantum systems (FCs and CENs)
and their signals operate. 'Classical' properties of systems emerge in two
cases. The first is 'large' spatial separations between FCs that allow them to
keep their local 'identity'. This can be used in reverse to define the
'largeness of the separations' by gauging the strength of the interactions
between two systems if any. The second is the 'large numbers effect' of many
bodies acting collectively as a single 'Newtonian' object. This is what we
usually mean by 'massive' objects whose behaviors can be described by
classical dynamics. Again 'massive' can be defined in reverse by identifying
the scale at which collective 'classical' properties of an object decouple
into the properties of the individual components.

\section{Complex Systems: Diffusion, Entropy, and Hierarchical Arrows of Time}

What happens when an ensemble of FCs act as a single $n$-body system? Lets
look first at gases. The thermodynamics of gases is well known for
'equilibrium' ensembles of atoms, molecules and gas-like particles such as
stars in galaxies and galaxies in clusters. Is the \textbf{diffusion }of a gas
a causal network? Yes, but what we shall call an 'open' on weakly 'wired' one
where the 'trajectories' of signals are probabilistic \cite{cert}. Diffusion
represents a special case in which the gas\ components forming the 'system'
act both as 'particles' and 'signals'. Their collective behaviors and motions
form a 'random' or 'open' network with a 'collective excitation' interpreted
as the 'temperature'. This is in contrast to the hard 'wiring' of
\ transistors and other components in integrated circuit chips whose electron
signal trajectories are well defined. Phase transitions correspond to CEs in
systems in which a population of FCs from a CEN. Prigogine recognized this
collective aspect to phase transitions:

\begin{quotation}
\textbf{''It is at this global level, at the level of populations, that the
symmetry between past and future is broken, and science can recognize the flow
of time'' }\emph{Ilya Prigogine} \cite{cert}
\end{quotation}

If we take this one step further we can see that these population 'arrows of
time' are the result of the interaction of the FCs in the population acting as
a new meso-FC system (CEN) with global CEs whose irreversible decay from an
excited state can be used to create the 'CEN Arrows of Time' he is observing
in the collective 'resonances' of complex systems from gases to the brain. At
this point the global behaviors of an 'open' or 'gas-like' system may be
described using probabilistic methods ignoring the 'identity' of the
individual FCs leading to a statistical formulation of classical mechanics.
One can 'define' an arrow time associated wit this global behavior, but this
is the scale at which irreversibility emerges. As we have seen above the QAT
associated with FCs and CENs gives us a glimpse of the fundamental source of
irreversible behavior at the microscopic level. This method does not apply for
'closed' or 'hard-wired' non-gas-like causal networks of FCs (CENs) supporting
collective excitations and behaviors. The methods of quantum computation
\cite{qcqi}, the method of collective excitations \cite{zim}, \cite{zago},
\cite{heyde}, and causal network theory are more appropriate for the
description of 'arrows of time' at plateaus of complexity (POCs) in many body
systems when one wants to preserve the quantum aspects of the FCs and yet
describe the global behaviors of CENs without resorting to '\textbf{coarse
graining}'.

\textbf{Entropy }production, $P(E)$, (as a function of the '\emph{energy of
reconfiguration}') by a Feynman Clock decaying from an excited state to a
ground state with an change in energy given by, $\Delta E_{\nu}=E^{\ast}%
-E_{0}=h\nu=hc/\lambda$ , provides another way of manufacturing 'time'. We
have the production of 'entropy' term \textbf{per unit 'time'}, $P(E)$, for a
non-isolated system interacting with its environment defined by:%

\begin{equation}
P(E)=\frac{dS_{FC}}{dt_{entropy}}%
\end{equation}

Solving for the 'lifetime' of this entropy production process and seeing that
this is just the same as the 'decay' lifetime for the 2-state FC above we have:%

\begin{equation}
\Delta t_{entropy}=\int\limits_{t(E_{0})}^{t(E^{\ast})}dt_{entropy}=%
%TCIMACRO{\dint \limits_{P(E_{0})}^{P(E\ast)}}%
%BeginExpansion
{\displaystyle\int\limits_{P(E_{0})}^{P(E\ast)}}
%EndExpansion
\frac{1}{P(E)}dS_{FC}=\frac{\hbar}{\Delta E_{\nu}}=\frac{\hbar}{\Gamma_{FC}%
}=\frac{\hbar}{\left|  \langle1,E_{0}\mid H_{E^{\ast}\rightarrow E_{0}}%
\mid0,E^{\ast}\rangle\right|  ^{2}}=\tau_{FC}%
\end{equation}

This allows us to make a correspondence between the QAT of a 2-state FC model
to the 'thermodynamic arrow of time' associated with the production of
'entropy' by the decay of an unstable state where:%

\begin{equation}
dS_{FC}=dS_{\Delta E_{\nu}}+dS_{\lambda}+dS_{VAC}%
\end{equation}

This is the total entropy change in the compound system of the FC interacting
with it's 'vacuum environment', via the creation of a 'signal' , $\lambda$.
The internal entropy change of the FC is $dS_{\Delta E_{\nu}}$, the entropy
transfer across the FC 'boundary' (e.g. nuclear decay with the production of a
gamma photon) is $dS_{\lambda}$, and the entropy change in the 'vacuum
environment' is $dS_{VAC}$. Perhaps the most important point here is that the
\emph{dimensional conversion of reconfiguration information} using
Planck's\emph{\ }constant into the 'lifetime', $\tau_{FC}$, of the FC (primary
information) \textbf{connects} the '\emph{time difference}', $\Delta
t_{entropy}$, (secondary information) of the 'constructed' thermodynamic arrow
of time 'map' between two configuration states of the FC. This includes the
flow of information from the FC to it's environment by creation of a photon
'signal' projected onto the vacuum energy density of space.

The superposition of the photon on the vacuum creates the 'classical'
component of the total system. This classical state of a 'free' photon
propagating in space is the basis of the 'classical arrow of time' (CAT)
associated with the source-to-sink 'electromagnetic arrow of time' in
classical and relativistic electrodynamics. We see here that construction of
macroscopic \textbf{hierarchical arrows of time} used to map configuration
transitions in complex systems are built from the flow of physical information
originating in the irreversible QAT of FCs (and CENs) in conjunction with the
'reversible' CATs associated with the 'free' signals in space that create
unstable FC and CEN systems via the fundamental particle interactions at the
microscopic scale.

\section{The Conversion of Information into 'Time'}

At this point we will use the techniques of quantum computation to show how
signals generated by the irreversible decay of an excited state of a FC can
transfer information via signals to FDs in causal networks. The creation of
the time label numbers we read off clocks requires an observing system to
process quantum sate information into the maps of the 'cause' and 'effect' for
our sensory data from our environments. Time is a label or number associated
with 'processed' signal information held in a detector or memory.

\section{'Computing' Time and the Creation of 'Moments'}

In summary we can see that the basic way to compute 'time' from the flow of
information between systems in the universe can be built on a simple two state
model. This model is really a 'time-independent' quantum computer in which
fundamental irreversible quantum processes coupled to reversible information
flow in the systems classical environment can be used to build complex
hierarchical systems through the phenomena of collective excitations occurring
at '\textbf{Plateaus of Complexity' (POCs).} These POCs may be cellular
membranes or boundary surfaces encapsulating causal networks and quantum
computers processing chemical and higher level information.

\section{Acknowledgments}

I wish to give my deep thanks to E. Keith Hege, Vladimir Petrov, Arkadi
Lipkine, Andre Bormanis, and Paola Zizzi for the encouragement and support
that has kept me going forward. My wife Toni and the rest of my extended
family have lost our 'time' together while I have pursued this challenge. I
thank them all for their patience during my struggle to discover the 'language
of time'.

\end{document}